\documentclass[submitting]{nst}

\usepackage{subfigure,dcolumn}
\usepackage{epstopdf}
\usepackage{mhchem}
\usepackage{upgreek}

\usepackage{float}
\usepackage{appendix}
\usepackage{threeparttable}
\usepackage{ulem}
\begin{document}

\title{Progress of photonuclear cross sections for medical radioisotope production at the SLEGS energy domain}
\thanks{Supported by the National Key R\&D Program of China (2022YFA1602401), the National Natural Science Foundation of China (Grants No. 11961141004, No. U1832211, No. 11922501, No. 12325506) and the National Basic Science Data Center `Medical Physics DataBase' (No.NBSDC-DB-23).}

\author{Xuan Pang}
\affiliation{School of Physics, Beihang University, Beijing 100191, China}

\author{Bao-Hua Sun}
\email[Corresponding author, ]{bhsun@buaa.edu.cn}
\affiliation{School of Physics, Beihang University, Beijing 100191, China}

\author{Li-Hua Zhu}
\email[Corresponding author, ]{zhulh@buaa.edu.cn}
\affiliation{School of Physics, Beihang University, Beijing 100191, China}

\author{Guang-Hong Lu}
\affiliation{School of Physics, Beihang University, Beijing 100191, China}

\author{Hong-Bo Zhou}
\affiliation{School of Physics, Beihang University, Beijing 100191, China}

\author{Dong Yang}
\affiliation{College of Physics, Jilin University, Changchun 130012, China}
\affiliation{National Basic Science Data Center, Beijing 100190, China}

\begin{abstract}
Photonuclear reactions using a laser Compton scattering (LCS) gamma source provide a new method for producing radioisotopes for medical applications. Compared with the conventional method, this method has the advantages of a high specific activity and less heat. Initiated by the Shanghai Laser Electron Gamma Source (SLEGS), we conducted a survey of potential photonuclear reactions, $(\upgamma,n)$, $(\upgamma,p)$, and $(\upgamma,\upgamma')$ whose cross-sections can be measured at SLEGS by summarizing the experimental progress. In general, the data are rare and occasionally inconsistent. Therefore, theoretical calculations are often used to evaluate the production of medical radioisotopes. Subsequently, we verified the model uncertainties of the widely used reaction code TALYS-1.96, using the experimental data of the \ce{^100Mo}$(\upgamma,n)$\ce{^99Mo}, \ce{^65Cu}$(\upgamma,n)$\ce{^64Cu}, and \ce{^68Zn}$(\upgamma,p)$\ce{^67Cu} reactions.
\end{abstract}

\keywords{Medical radioisotope, Photonuclear reaction, LCS, Cross section}

\maketitle

\section{Introduction}\label{sec.I}
Radioisotopes are widely used for the diagnosis and therapy of different diseases owing to their nuclear-physical properties~\cite{b1,b2}. Diagnostic radioisotopes can provide functional and metabolic information for the early treatment of diseased regions that have not yet undergone morphological or structural changes. Currently, positron emission tomography (PET)~\cite{b3} and single-photon emission computed tomography (SPECT)~\cite{b4} are the two major diagnostic techniques. Short-lived $\upbeta^+$-emitting radioisotopes are often used for PET. Typical examples include \ce{^11C}, \ce{^13N}, \ce{^15O}, and \ce{^18F}, which have half-lives of 20, 10, 2, and 110 min, respectively. For SPECT, $\upgamma$-ray emitting radioisotopes are frequently used, of which \ce{^{99m}Tc} with a half-life of 6 \si{h}~\cite{b5} is the most common radioisotope tracer. Therapeutic radioisotopes can be combined with targeted drugs to achieve the precise removal of small diseased regions without excessive doses to normal tissues. Therefore, radioisotopes that emit low-range highly ionizing radiation are of significant interest. The $\upbeta^-$-particle emitting radioisotopes (\emph{e.g.} \ce{^32P}, \ce{^89Sr}, \ce{^90Y}, \ce{^131I}, \ce{^177Lu}, and \ce{^188Re}), Auger electron cascades (\emph{e.g.} \ce{^103Pd} and \ce{^125I}) and $\upalpha$-particle emitting radioisotopes (\emph{e.g.} \ce{^211At}, \ce{^212Bi}, \ce{^223Ra}, \ce{^225Ac}, and \ce{^227Th}), with a high linear energy transfer in tissue, are suitable for therapy.

More than 40 million nuclear medicine procedures are performed annually, and the demand for radioisotopes is increasing by 5\% annually~\cite{b6}. Currently, the production of medical radioisotopes relies primarily on nuclear reactors and cyclotrons. Thermal neutron-induced fission produces neutron-rich radioisotopes in nuclear reactors. The advantage of this method is the possibility of producing high levels of total and specific activities. However, the production of desired radioisotopes is accompanied by a considerable amount of long-lived radioactive waste, which raises numerous safety and security concerns. In addition, many nuclear reactors used for medical radioisotope production are more than 50 years old and may be shut down in the near future~\cite{b7}. For example, a vast majority of reactors producing \ce{^99Mo} are expected to shut down by 2030. Compared with reactors, cyclotrons typically produce neutron-deficient radioisotopes by charged particle reactions accompanied by less radioactive waste. In addition, they have the advantage of being more compact, allowing their placement near hospitals, and making them a useful tool for producing radioisotopes with short half-lives that range from several minutes to hours.

Another promising method for radioisotope production involves photonuclear reactions, mainly including $(\upgamma,n)$, $(\upgamma,p)$, and $(\upgamma,\upgamma')$. This is considered an alternative to radioisotope production in reactors and cyclotrons or the only method of production for certain radioisotopes. $(\upgamma,n)$ reactions can produce $\upbeta^+$-emitting radioisotopes for PET imaging, such as \ce{^11C}, \ce{^13N}, \ce{^15O}, and \ce{^18F}. These radioisotopes are produced by cyclotron-based $(p,n)$ or $(p,\upgamma)$ reactions. $(\upgamma,p)$ reactions are suitable for producing $\upbeta^-$-emitting radioisotopes that are currently mostly produced in reactors. High-intensity $\upgamma$ beams are required to obtain adequate yield using this method. With the development of laser Compton scattering (LCS) gamma sources, the production of radioisotopes via photonuclear reactions has attracted considerable attention~\cite{b8,b9,b10,b11,b12,b13,b14,b15,b16,b17}.

In contrast to conventional bremsstrahlung gamma sources based on electron linear accelerators, LCS gamma sources have the advantages of a high photon flux and excellent monochromaticity. The advantages of photonuclear reactions using LCS gamma sources for radioisotope production are as follows:

\begin{itemize}
    \item Differing from bremsstrahlung gamma sources, the LCS gamma sources significantly reduce the heat per useful reaction rate as its $\upgamma$-rays in the energy range of interest are not accompanied by an intense low-energy tail. Moreover, the LCS gamma sources have the ability to selectively tune photons to energies of interest. A high specific activity can be achieved by matching the photon energy to the giant dipole resonance (GDR) peak.
    \item Differing from the charged-particle induced reactions, the photonuclear reactions generate less heat as their energy deposition in targets via gamma-matter interactions is significantly smaller. Thus, the target can be thicker and requires considerably less cooling~\cite{b9}. Moreover, the photonuclear reactions have the capability to simultaneously irradiate multiple targets. In this manner, it can maximize the utilization of $\upgamma$ beams and produce multiple radioisotopes simultaneously~\cite{b8}.
\end{itemize}

The Shanghai Laser Electron Gamma Source (SLEGS) employs the LCS technique. It can generate $\upgamma$ beams ranging from 0.25–21.7 \si{MeV} with a full-spectrum flux of $10^{5}$–$10^{7}$ \si{s^{-1}}~\cite{b18} and the best possible bandwidth of $\upgamma$ beams of 5\%–15\% after passing through a dual collimation system~\cite{b19}. Using collimation technology, the size of an $\upgamma$ beam can be continuously adjusted to within $\Phi$25 \si{mm}~\cite{b20}. The monochromaticity and high intensity of $\upgamma$ beams combined with detector spectrometers can be used to measure the photonuclear reaction cross-section ~\cite{b21,b22,b23,b24}. One of the main topics of SLEGS is the measurement of the key photonuclear reaction cross-section relevant for medical radioisotope investigations, providing crucial nuclear data for the photonuclear method.

In this study, we aim to estimate the production of medical radioisotopes in the energy range of 0.25–21.7 \si{MeV} in the SLEGS domain. The remainder of this paper is organized as follows. In Sect.~\ref{sec.II}, we summarize the radioisotopes that can be produced via $(\upgamma,n)$, $(\upgamma,p)$, and $(\upgamma,\upgamma')$ reactions and discuss their experimental feasibility. Owing to the scarcity of cross-sectional data regarding photonuclear reactions, theoretical calculations are often used to assess the yields of medical radioisotopes. In Sect.~\ref{sec.III}, the cross-sectional data of the \ce{^100Mo}$(\upgamma,n)$, \ce{^65Cu}$(\upgamma,n)$, and \ce{^68Zn}$(\upgamma,p)$ reactions in the SLEGS energy region were investigated to produce \ce{^99Mo}, \ce{^64Cu}, and \ce{^67Cu} radioisotopes. Finally, we provide a summary in Sect. ~\ref{sec.V}.

\section{Medical radioisotopes produced in photonuclear reactions}\label{sec.II}
In this section, we discuss potential $(\upgamma,n)$, $(\upgamma,p)$, and $(\upgamma,\upgamma')$ reactions for the production of medical radioisotopes. Crucial considerations include the natural abundance of the target isotopes, half-life of the radioisotopes, and available experimental data. Generally, it is preferable to have a target isotope with a dominant abundance (\emph{e.g., }, greater than 10\%), and a suitable half-life (several tens of minutes to days) for the produced radioisotope. However, the lack of experimental data may limit the medical applications of radioisotopes.

Different methods are available for measuring photonuclear reactions, including in-beam and offline measurements. Four $\pi$ \ce{^3He} neutron detection arrays, charged particle detectors, and nuclear resonance fluorescence spectrometers are commonly used for in-beam measurements of $(\upgamma,n)$, $(\upgamma,p)$, and $(\upgamma,\upgamma')$ reactions. Offline measurement is suitable for determining all three types of photonuclear reactions only in cases where the residual decay has an unambiguous $\upgamma$-ray and the lifetime of the residual nucleus is within a range of minutes to hours.

\subsection{$(\upgamma,n)$ reactions}
Radioisotopes can be most efficiently produced by exciting the target nuclide into a GDR using photons. The GDR is characterized by a large peak cross-section and broad width, resulting in a large integral cross-section. The shape of the GDR follows a Lorentzian distribution with a spreading width of approximately 5 \si{MeV}~\cite{b25,b26,b27}. Generally, a considerable portion of the total photonuclear cross-section is known to originate from the $(\upgamma,n)$ reactions, which increases as the charge number of the target nuclide increases, making the following chemical separation easier. For example, the \ce{^100Mo}$(\upgamma,n)$ reaction accounts for more than 70\% of the total photonuclear reactions within the energy range of 8.4–15 \si{MeV}. Therefore, medium- and heavy-mass nuclei are ideal for production via $(\upgamma,n)$ reactions~\cite{b28}.

Table~\ref{table1} summarizes the 15 medical radioisotopes produced by $(\upgamma,n)$ reactions~\cite{b8,b14,b29,b30,b31}. Excluding \ce{^100Mo}, the natural abundance of these reaction targets exceeds 10\%. A higher natural abundance results in a higher yield and lower impurity content. The seventh column indicates the current status of the EXFOR database~\cite{b32}. The experimental data for only nine radioisotopes are presented in the Appendix. Among these, the \ce{^19F}$(\upgamma,n)$ reaction has only one data set that was measured in the 1960s. The data for the \ce{^90Zr}$(\upgamma,n)$, \ce{^100Mo}$(\upgamma,n)$, and \ce{^187Re}$(\upgamma,n)$ reactions do not cover the entire GDR energy region. There is a significant discrepancy between the two sets of data for the \ce{^65Cu}$(\upgamma,n)$ reaction: It is possible to determine the relevant cross-sections at SLEGS.

\begin{table*}[!htb]
\caption{Medical radioisotope $^{A}X$ production via the $^{A+1}X(\upgamma,n)^{A}X$ reactions~\cite{b8,b14,b29,b30,b31}. The natural abundance of $^{A+1}X$ is provided. The produced medical radioisotopes $^{A}X$ are characterized by the half-lives $T_{1/2}$ and decay modes. Here, $\%\upvarepsilon$ represents the probabilities of nuclear decay via electron capture or $\upbeta^+$ decay. $\%\upbeta^-$ represents the probabilities of $\upbeta^-$ decay. $E_{th}$ represents the threshold energy of the reaction. The seventh column indicates whether the $^{A+1}X(\upgamma,n)^{A}X$ reactions have experimental data in the EXFOR database~\cite{b32}. $a$ and $b$ indicate that the radioisotope can be studied by in-beam and off-line measurements, respectively. The last column presents the probable $\upgamma$-rays of interest in the offline measurement. The data are obtained from Ref.~\cite{b33}.}
\label{table1}
\begin{tabular*}{\linewidth} {@{\extracolsep{\fill} } cccccccccc}
\toprule
$^{A+1}X$    & Nat. abu. (\%) &         $^{A}X$           & $T_{1/2}$      &      Decay mode        &  $E_{th}$ (\si{MeV})  & EXFOR data & Medical application & Exp. method & $\upgamma$-ray (\si{keV}) \\
\midrule
\ce{^12C}    & 98.93          &      \ce{^11C}            & 20.364 \si{m}  & $\%\upvarepsilon=100$  &         18.7          &     Yes    &         PET         & $a$;$b$ & 511 \\
\ce{^14N}    & 99.63          &      \ce{^13N}            & 9.965 \si{m}   & $\%\upvarepsilon=100$  &         10.5          &     Yes    &         PET         & $a$;$b$ & 511 \\
\ce{^16O}    & 99.757         &      \ce{^15O}            & 122.24 \si{s}  & $\%\upvarepsilon=100$  &         15.6          &     Yes    &         PET         &  $a$  &  -  \\
\ce{^19F}    & 100            &      \ce{^18F}            & 109.77 \si{m}  & $\%\upvarepsilon=100$  &         10.4          &     Yes    &         PET         & $a$;$b$ & 511\\
\ce{^45Sc}   & 100            &      \ce{^44Sc}           & 3.97 \si{h}    & $\%\upvarepsilon=100$  &         11.3          &     No     &         PET         & $a$;$b$ & 511;1157 \\
\ce{^63Cu}   & 69.15          &      \ce{^62Cu}           & 9.673 \si{m}   & $\%\upvarepsilon=100$  &         10.8          &     Yes    &         PET         & $a$;$b$ & 511 \\
\ce{^65Cu}   & 30.85          &      \ce{^64Cu}           & 12.7 \si{h}    & $\%\upvarepsilon=61.5$ &         9.91          &     Yes    &   PET;Radiotherapy  & $a$;$b$ & 511 \\
             &                &                           &                & $\%\upbeta^-=38.5$     &                       &            &                     &     &   \\
\ce{^69Ga}   & 60.108         &      \ce{^68Ga}           & 67.71 \si{m}   & $\%\upvarepsilon=100$  &         10.3          &     No     &         PET         & $a$;$b$ & 511;1077 \\
\ce{^90Zr}   & 51.45          &      \ce{^89Zr}           & 78.41 \si{h}   & $\%\upvarepsilon=100$  &         11.9          &     Yes    &         PET         &  $a$  & - \\
\ce{^100Mo}  & 9.82           &      \ce{^99Mo}           & 65.976 \si{h}  & $\%\upbeta^-=100$      &         8.29          &     Yes    &        SPECT        & $a$;$b$ & 740;778 \\
\ce{^104Pd}  & 11.14          &      \ce{^103Pd}         &  17.0 \si{d}   & $\%\upvarepsilon=100$  &         10.0          &     No     &     Radiotherapy    &  $a$  & - \\
\ce{^166Er}  & 33.503         &      \ce{^165Er}          & 10.36 \si{h}   & $\%\upvarepsilon=100$  &         8.47          &     No     &     Radiotherapy    &  $a$  & - \\
\ce{^170Er}  & 14.91          &      \ce{^169Er}          &  9.4 \si{d}    & $\%\upbeta^-=100$      &         7.25          &     No     &     Radiotherapy    &  $a$  & - \\
\ce{^187Re}  & 62.60          &      \ce{^186Re}          &  3.7 \si{d}    & $\%\upbeta^-=92.53$    &         7.36          &     Yes    &  SPECT;Radiotherapy &  $a$  & - \\
             &                &                           &                & $\%\upvarepsilon=7.47$ &                       &            &                     &     &   \\
\ce{^193Ir}  & 62.70          &      \ce{^192Ir}          & 73.8 \si{d}    & $\%\upbeta^-=95.24$    &         7.77          &     No     &     Radiotherapy    &  $a$  & - \\
             &                &                           &                & $\%\upvarepsilon=4.76$ &                       &            &                     &     &   \\
\bottomrule   
\end{tabular*}
\end{table*}

\subsection{$(\upgamma,p)$ reactions}
Despite the nucleus being excited beyond the proton separation energy of the photons, it does not necessarily lose a proton. Only for excitations well beyond the proton separation energy, the proton can acquire sufficient kinetic energy to effectively cross the Coulomb barrier. However, in these cases, the excitation energy typically exceeds the separation energies of one or two neutrons. Therefore, the neutron emission channel competes with the proton emission channel. For light and certain medium-mass nuclei, the cross-sections of the $(\upgamma,p)$ reactions are comparable to and occasionally exceed those of the $(\upgamma,n)$ reactions, owing to the shell structure of the nuclei. Certain heavier $\upbeta^-$ emitters for radionuclide therapy can also be produced by $(\upgamma,p)$ reactions, but the increasing Coulomb barrier leads to the production of small cross-sections. $(\upgamma,p)$ reactions result in the daughter and parent isotopes being chemically different. Several advanced chemical-separation techniques have been developed for this purpose. For example, a simple and reproducible three-ion exchange matrix approach was developed to separate \ce{^67Cu} from \ce{^68Zn}~\cite{b34}. Separation techniques exist for other pairs, such as Ti/Sc, Zr/Y, and Hf/Lu~\cite{b35,b36,b37}. 

The potential $(\upgamma,p)$ reactions used for nuclear medicine are listed in Table~\ref{table2}. However, the experimental data regarding this topic are scarce. Only limited data for \ce{^43K}, \ce{^67Cu}, and \ce{^177Lu} has been obtained by using the bremsstrahlung gamma source, as detailed in the Appendix. No error bars are shown for the \ce{^43K} data, whereas relatively large errors ranging from typically 10\% to 100\% for \ce{^67Cu} and \ce{^177Lu} have been demonstrated.

\begin{table*}[!htb]
\caption{Same as Table~\ref{table1} but for medical radioisotopes $^{A}X$ production via the $^{A+1}Y(\upgamma,p)^{A}X$ reactions~\cite{b8,b38,b39,b40}. Nat. abu. indicate the natural abundance of $^{A+1}Y$. The seventh column indicates whether experimental data regarding the $^{A+1}Y(\upgamma,p)^{A}X$ reactions is available in the EXFOR database~\cite{b32}.}
\label{table2}
\begin{tabular*}{\linewidth} {@{\extracolsep{\fill} } cccccccccc}
\toprule
$^{A+1}Y$    & Nat. abu. (\%) &         $^{A}X$           & $T_{1/2}$      &     Decay mode         &  $E_{th}$ (\si{MeV})  & EXFOR data & Medical application & Exp. method & $\upgamma$-ray (\si{keV}) \\
\midrule
\ce{^44Ca}   & 2.09           &      \ce{^43K}            & 22.3 \si{h}    & $\%\upbeta^-=100$      &         12.1          &     Yes    &    Radiotherapy     & $a$;$b$ & 373;617 \\
\ce{^48Ti}   & 73.72          &      \ce{^47Sc}           & 3.35 \si{d}    & $\%\upbeta^-=100$      &         11.4          &     No     &    Radiotherapy     &  $a$  & - \\
\ce{^58Ni}   & 68.077         &      \ce{^57Co}           & 271.74 \si{d}  & $\%\upvarepsilon=100$  &         8.17          &     No     &    Radiotherapy     &  $a$  & - \\
\ce{^68Zn}   & 18.45          &      \ce{^67Cu}           & 61.83 \si{h}   & $\%\upbeta^-=100$      &         9.97          &     Yes    &    Radiotherapy     & $a$;$b$ & 861;2502 \\
\ce{^91Zr}   & 11.22          &      \ce{^90Y}            & 64.053 \si{h}  & $\%\upbeta^-=100$      &         8.68          &     No     &    Radiotherapy     &  $a$  & - \\
\ce{^106Pd}  & 27.33          &      \ce{^105Rh}          & 35.36 \si{h}   & $\%\upbeta^-=100$      &         9.34          &     No     &    Radiotherapy     & $a$;$b$ & 319 \\
\ce{^112Sn}  & 0.97           &      \ce{^111In}          & 2.805 \si{d}   & $\%\upvarepsilon=100$  &         7.55          &     No     &  SPECT;Radiotherapy &  $a$  & - \\
\ce{^132Xe}  & 26.9           &      \ce{^131I}           & 8.025 \si{d}   & $\%\upbeta^-=100$      &         9.12          &     No     &    Radiotherapy     &  $a$  & - \\
\ce{^162Dy}  & 25.475         &      \ce{^161Tb}          & 6.89 \si{d}    & $\%\upbeta^-=100$      &         8.00          &     No     &  SPECT;Radiotherapy &  $a$  & - \\
\ce{^167Er}  & 22.869         &      \ce{^166Ho}          & 26.824 \si{h}  & $\%\upbeta^-=100$      &         7.50          &     No     &  SPECT;Radiotherapy &  $a$  & - \\
\ce{^178Hf}  & 27.28          &      \ce{^177Lu}          & 6.647 \si{d}   & $\%\upbeta^-=100$      &         7.34          &     Yes    &  SPECT;Radiotherapy & $a$;$b$ & 208 \\
\ce{^189Os}  & 16.15          &      \ce{^188Re}          &  17 \si{h}     & $\%\upbeta^-=100$      &         7.25          &     No     &  SPECT;Radiotherapy &  $a$  & - \\
\bottomrule
\end{tabular*}
\end{table*}

\subsection{$(\upgamma,\upgamma')$ reactions}
A nuclear isomer is the metastable state of an atomic nucleus, in which one or more nucleons (protons or neutrons) occupy higher energy levels than the ground state of the same nucleus. The lifetime and excitation energy are two important properties of nuclear isomers. The excitation energy between the isomer and ground state is characteristic and can be used to identify nuclides. Nuclear isomers can be deexcited by the emission of $\upgamma$-rays and/or the conversion of electrons to the ground state. The $\upgamma$-emitting isomers can be used as radioactive labels for SPECT imaging. Nuclear isomers emitting low-energy Auger electrons are potential radioisotopes for targeted therapy.

Conventional production methods, such as $(n,\upgamma)$ reactions, have relatively low yields because the dominant part of the production proceeds directly to the nuclear ground state with a spin closer to that of the target isotope. Using monochromatic and small-bandwidth LCS photons, transitions from a stable or long-lived nuclear ground state to higher energy levels can be selectively excited. Such levels serve as gateway states, which then partially decay to the isomeric state directly or via a cascade. The $(\upgamma,\upgamma')$ reaction is equivalent to storing the energy of an incident photon in an isotope that acts as a container.

Table~\ref{table3} lists the six isomers that can be produced by $(\upgamma,\upgamma')$ reactions for medical applications. Excluding \ce{^{117m}Sn} and \ce{^{176m}Lu}, the experimental data for the other radioisotopes are detailed in the Appendix. The dataset of \ce{^{103m}Rh} did not have error bars, and two different results were reported from 6 to 23 \si{MeV}. For \ce{^{113m}In}, two cross-sectional datasets differ by more than 300 times at 8 \si{MeV}. For \ce{^{115m}In}, there were significant discrepancies between multiple datasets. The data of \ce{^{195m}Pt} have a large relative error as high as 100\% at 3.5, 7.5, and 8 \si{MeV}.

\begin{table*}[!htb]
\caption{Same as Table~\ref{table1} but for medical isomers $^{Am}X$ production via the $^{A}X(\upgamma,\upgamma')^{Am}X$ reactions~\cite{b8,b13}. The isomeric states $^{Am}X$ are characterized by the energy $E_m$, half-lives $T^m_{1/2}$, and decay modes. Here, $\%IT$ represents the probabilities of nuclear decay via isomeric transitions. $E_{th}$ equals $E_m$. The seventh column indicates whether the $^{A}X(\upgamma,\upgamma')^{Am}X$ reactions have experimental data in the EXFOR database~\cite{b32}. $^{A}X(\upgamma,\upgamma')^{Am}X$ reactions can be investigated by off-line measurements where the $\upgamma$-ray energy of interest is equal to $E_m$.}
\label{table3}
\begin{tabular*}{\linewidth} {@{\extracolsep{\fill} } cccccccc}
\toprule
$^{A}X$      &  Nat. abu. (\%)  &       $^{Am}X$      &  $T^m_{1/2}$     &  $E_m$/$E_{th}$ (keV) &    Decay mode          & EXFOR data & Medical application \\
\midrule
\ce{^103Rh}  &      100         &    \ce{^{103m}Rh}   &   56.114 \si{m}  &         39.8          & $\%IT=100$             &     Yes    &     Radiotherapy    \\
\ce{^113In}  &      4.29        &    \ce{^{113m}In}   &   99.476 \si{m}  &         391.7         & $\%IT=100$             &     Yes    &  SPECT;Radiotherapy \\
\ce{^115In}  &      95.71       &    \ce{^{115m}In}   &   4.486 \si{h}   &         336.2         & $\%IT=95$              &     Yes    &  SPECT;Radiotherapy \\
             &                  &                     &                  &                       & $\%\upbeta^-=5$        &            &                     \\
\ce{^117Sn}  &      7.68        &    \ce{^{117m}Sn}   &   13.76 \si{d}   &         314.6         & $\%IT=100$             &     No     &     Radiotherapy    \\
\ce{^176Lu}  &      2.599       &    \ce{^{176m}Lu}   &   3.664 \si{h}   &         122.8         & $\%\upbeta^-=99.90$    &     No     &     Radiotherapy    \\
             &                  &                     &                  &                       & $\%\upvarepsilon=0.09$ &            &                     \\
\ce{^195Pt}  &      33.78       &    \ce{^{195m}Pt}   &   4.01 \si{d}    &         259.3         & $\%IT=100$             &     Yes    &     Radiotherapy    \\
\bottomrule
\end{tabular*}
\end{table*}

\section{Evaluation of cross-sections for \ce{^99Mo}, \ce{^64Cu}, and \ce{^67Cu} production}\label{sec.III}
The cross-section data were used to calculate the expected yield of the radioisotopes for a given thickness and enrichment of the target material, and to determine the optimum energy range for the production of the desired radioisotope and the level of radioisotopic impurities. However, the slow development of gamma sources and small cross-sections of photonuclear reactions resulted in a relatively small number of experimental studies regarding this cross-sectional data. There is often scattering between different experimental datasets. As a result, the cross-sections predicted by the theoretical models play a key role in the production of medical radioisotopes.

The TALYS code~\cite{b41} offers a unified approach for calculating nuclear reactions that involve neutrons, photons, protons, deuterons, tritons, \ce{^3He}, and $\upalpha$ particles in the \si{keV}-200 \si{MeV} energy range and for target nuclides with masses of 5 and 339. The code outputs a set of reaction data, \emph{ for example } cross-sections, energy spectra, and angular distributions of the emitted particles, \emph{etc.}. Numerous studies have tested the TALYS code and demonstrated that it has reliable predictive power in terms of the nuclear reaction calculations~\cite{b42,b43,b44}. In the TALYS code, the decay of the compound nuclear state from the photonuclear reaction was treated using the Hauser-Fenshbach (HF) statistical model~\cite{b45}. The nuclear-level density and $\upgamma$ strength function were the main input parameters. In this study, we adapted the TALYS-1.96 code. It implemented six different nuclear level density (NLD) models and nine different models of $\upgamma$ strength functions ($\upgamma$SF), as shown in Table~\ref{table4}. The NLD and $\upgamma$SF models employ phenomenological and microscopic approaches, respectively.

In this section, we select the \ce{^99Mo}, \ce{^64Cu}, and \ce{^67Cu} radioisotopes as examples to compare the model calculations with the experimental data. \ce{^{99m}Tc} decayed by \ce{^99Mo} is the most frequently used radioisotope in nuclear medicine. \ce{^68Zn} is of interest owing to its applications in the production of the medical radioisotope \ce{^67Cu}; \ce{^67Cu} and \ce{^64Cu} can form a "matched pair"~\cite{b46,b47}. Therapeutic \ce{^67Cu}, along with positron-emitting \ce{^64Cu}, can measure the uptake kinetics in an organ of a patient by PET imaging, allowing for a precise dosimetric calculation. We used the default values of NLD and the $\upgamma$SF models in the following calculations.

\begin{table*}[!htb]
\caption{The descriptions of NLD and $\upgamma$SF models in the TALYS-1.96 code}
\label{table4}
\begin{tabular*}{\linewidth} {@{\extracolsep{\fill} } cc}
\toprule
Input parameter      &                      Detailed description                              \\
\midrule
ldmodel 1 (default)  &            Constant Temperature + Fermi gas model                      \\
ldmodel 2            &                 Back-shifted Fermi gas model                           \\
ldmodel 3            &                 Generalised Superfluid model                           \\
ldmodel 4            &  Skyrme-Hartree-Fock-Bogoluybov level densities from numerical tables  \\
ldmodel 5            &   Gogny-Hartree-Fock-Bogoluybov (GHFB) level densities from numerical tables  \\
ldmodel 6            &      Temperature-dependent GHFB level densities from numerical tables  \\
strength 1 (default for incident neutrons)        &    Kopecky-Uhl generalized Lorentzian     \\
strength 2 (default for other incident particles) &        Brink-Axel Lorentzian        \\
strength 3           &         Hartree-Fock + Bardeen-Cooper-Schrieffer (BCS) tables    \\
strength 4           &                  Hartree-Fock-Bogoliubov (HFB) tables            \\
strength 5           &                        Goriely's hybrid model                    \\
strength 6           &              Goriely Temperature-dependent HFB model             \\
strength 7           &     Temperature-dependent Relativistic Mean Field (RMF) model    \\
strength 8           & Gogny D1M HFB + Quasiparticle Random Phase Approximation (QRPA)  \\
strength 9           &           Simplified Modified Lorentzian (SMLO) model            \\
\bottomrule
\end{tabular*}
\end{table*}

\subsection{\ce{^99Mo}/\ce{^{99m}Tc}}
\begin{figure}[!htb]
\includegraphics[width=\hsize]{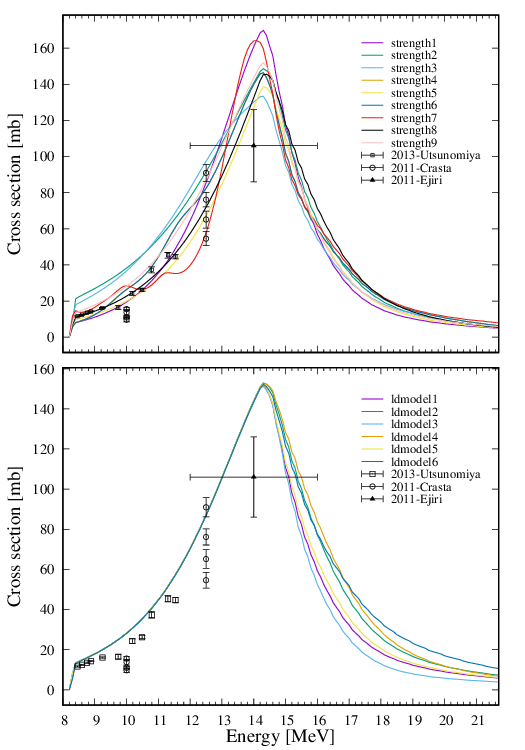}
\caption{(Color online) \ce{^100Mo}$(\upgamma,n)$\ce{^99Mo} reaction cross sections.}
\label{fig1}
\end{figure}

With half-lives of 6 \si{h} and 140 \si{keV} $\upgamma$-rays, \ce{^{99m}Tc} is nearly ideal for SPECT imaging. The most common method to obtain \ce{^{99m}Tc} is by elution from \ce{^99Mo} generators. The \ce{^100Mo}$(\upgamma,n)$ reaction has a sufficient potential to produce \ce{^99Mo}. The excitation functions are presented in Fig.~\ref{fig1}, which covers the entire GDR energy range. The cross-sections of \ce{^100Mo}$(\upgamma,n)$\ce{^99Mo} were experimentally measured by Utsunomiya et al.~\cite{b48}, Crasta et al.~\cite{b49}, and Ejiri et al.~\cite{b14}. The gamma sources used in the experiments were LCS and bremsstrahlung. All the experimental results were distributed in the low-energy region of the GDR and did not exceed the peak. Fig.~\ref{fig1} demonstrates that the shape and value of the excitation function are highly sensitive to the choice of the $\upgamma$SF model, particularly at the peak position of GDR. Excluding Strength 2 and 3, the remaining $\upgamma$SF models yield relatively similar results that are close to the experimental results. Different $\upgamma$SF models produce similar predictions for energies above the GDR peak. Overall, the cross-sections estimated using Strength 8 achieve the best agreement with the experimental data. As shown in Fig.~\ref{fig1}, the calculation results of all the NLD models below the peak are consistent but significantly higher than the experimental data, whereas those above the peak are significantly different. The data predicted by the models at energies above 14 \si{MeV} are highly valuable. New experiments can be conducted based on SLEGS to verify the model predictions.

\subsection{\ce{^64Cu}}
\begin{figure}[!htb]
\includegraphics[width=\hsize]{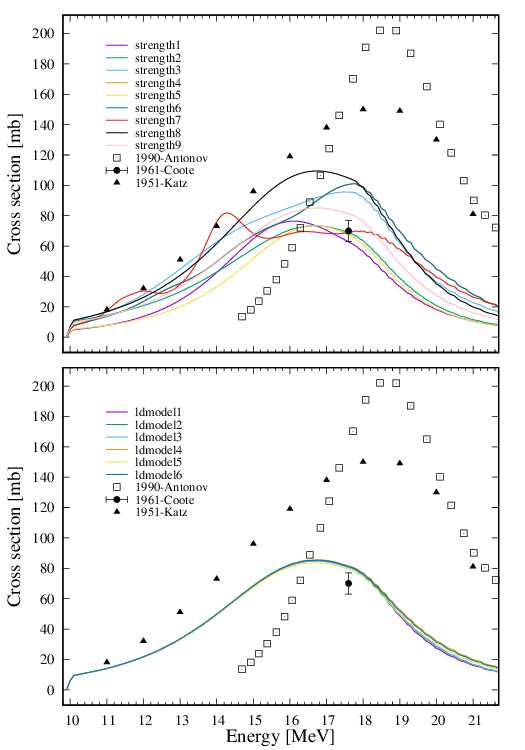}
\caption{(Color online) \ce{^65Cu}$(\upgamma,n)$\ce{^64Cu} reaction cross sections.}
\label{fig2}
\end{figure}

As listed in Table~\ref{table1}, the decay characteristics of \ce{^64Cu} render it useful for nuclear medicine~\cite{b50}. It combines PET diagnostic capabilities with those of radiotherapy, with average electron emissions of 190 \si{keV}. Currently, \ce{^64Cu} is mainly produced by small cyclotrons through \ce{^64Ni}$(p,n)$ reactions~\cite{b51,b52,b53,b54}. Alternative production using \ce{^65Cu}$(\upgamma,n)$ does not require rare or expensive \ce{^64Ni} targets and simplifies the chemical separation step. The experimental data for the cross-sections of the \ce{^65Cu}$(\upgamma,n)$\ce{^64Cu} reaction are presented in Fig.~\ref{fig2} along with the TALYS. The measurements by Katz et al.~\cite{b55} and Antonov et al.~\cite{b56} were performed using bremsstrahlung gamma sources. However, they differed significantly in terms of their shape and value. Coote et al.~\cite{b57} obtained a single cross-section at 17.6 \si{MeV} using monochromatic $\upgamma$-rays from the \ce{^7Li}$(p,\upgamma)$\ce{^8Be} reaction. The cross-section is consistent with the calculation but lower than other experimental data by 80 to 120 \si{mb}. The cross-sections of \ce{^65Cu}$(\upgamma,n)$\ce{^64Cu} can significantly vary with the choice of the $\upgamma$SF models but not the NLD models. However, the discrepancies between the experiments and models cannot be compensated for by varying the $\upgamma$SF models. Future experiments regarding the \ce{^65Cu} photoneutron reaction using a monochromatic LCS gamma source will be useful for resolving these discrepancies and provide assurance for its medical applications.

\subsection{\ce{^67Cu}}
\ce{^67Cu} is a promising $\upbeta^-$-particle emitter with an average energy of 141 \si{keV} for the targeted radiotherapies. The size of these particles in the tissue was of the same order as the cell diameter. This reduces the unwanted dose burden on patients ~\cite{b58}. With a half-life of 61.83 h and low energy $\upgamma$ emissions (91.266 \si{keV}, 7\%; 93.311 \si{keV}, 16.1\%; 184.577 \si{keV}, 48.7\%), \ce{^67Cu} can provide a long therapeutic effect. \ce{^67Cu} and its stable daughter \ce{^67Zn} are nontoxic to the body, and both Cu and Zn are prevalent trace elements in the body. Along with the PET imaging radioisotope \ce{^64Cu}, it forms a ``matched pair.'' However, the widespread clinical use of \ce{^67Cu}-based radiopharmaceuticals has been limited by their availability, quantity, and quality. Experiments~\cite{b59,b60,b61,b62,b63} have shown that the \ce{^68Zn}$(\upgamma,p)$ reaction has the potential to produce sufficient quantities of \ce{^67Cu} with a sufficient purity for medical use. Moreover, various Cu and Zn separation methods have been employed in radiochemical processing~\cite{b34,b64,b65,b66}.

\begin{figure}[!htb]
\includegraphics[width=\hsize]{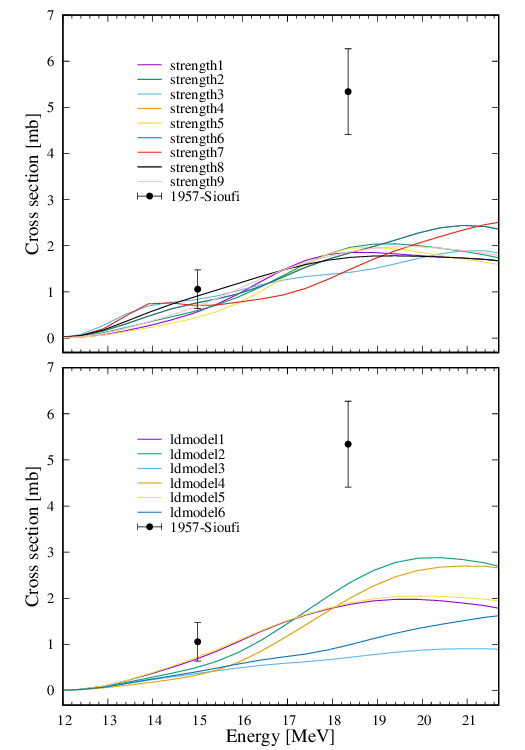}
\caption{(Color online) \ce{^68Zn}$(\upgamma,p)$\ce{^67Cu} reaction cross sections.}
\label{fig3}
\end{figure}

Figure~\ref{fig3} presents the excitation function of the \ce{^68Zn}$(\upgamma,p)$\ce{^67Cu} reaction. Experimental data regarding this reaction are scarce, with only one group of two points in the energy region of SLEGS~\cite{b67}. The data are significantly inconsistent with any calculation and hardly constrain NLD and $\upgamma$SF models. In contrast to the $(\upgamma,n)$ reaction of several hundred \si{mb}, the $(\upgamma,p)$ reaction is typically several \si{mb} in the medium-heavy nuclei. Additionally, protons can easily stop at the target and are difficult to detect. If the product nuclei of the $(\upgamma,p)$ reaction are unstable and have a moderate half-life, the cross-sections can be determined by offline measurements of the characteristic $\upgamma$-rays during decay. For the \ce{^68Zn}$(\upgamma,p)$ reaction, an offline measurement technique can be employed to determine its cross-sections.

\section{Summary} \label{sec.V}
In this study, we summarize the medical radioisotopes that can be produced by $(\upgamma,n)$, $(\upgamma,p)$, and $(\upgamma,\upgamma')$ reactions, along with the experimental data for these production pathways. We investigated the photonuclear reaction cross-sections for the production of \ce{^99Mo}, \ce{^64Cu}, and \ce{^67Cu}. The experimental data for \ce{^99Mo} and \ce{^67Cu} did not cover the entire GDR energy region, whereas the data for \ce{^64Cu} exhibited a significant discrepancy. To better constrain the NLD and $\upgamma$SF models in TALYS, additional experimental measurements at SLEGS in the energy region of GDR are necessary in the future.

To compare the productivity of the photonuclear method with that of traditional methods, we consider the production of \ce{^99Mo}/\ce{^{99m}Tc} as an example. The highest flux currently available for SLEGS is $10^7$ \si{s^{-1}}. We adopted the calculated results by the TALYS-1.96 code using the Strength 8 model for the yield calculations, which provided a maximum of 145 \si{mb} of the \ce{^100Mo}$(\upgamma,n)$ reaction cross-section at the incident photon energy of 14.4 \si{MeV}. After irradiating a \ce{^100MoO_3} target with a radius of 2 \si{mm} and a thickness of 1 \si{cm} for 5 times the half-life of \ce{^99Mo}, the saturation specific activity of \ce{^99Mo} and \ce{^{99m}Tc} can reach 4.76 and 4.74 \si{µCi/g}, respectively. The use of stacked targets can further enhance production. When the energy regions corresponding to the maximum cross-sections of the targeted nuclear reactions are close, multiple radioisotopes can be simultaneously produced. For example, the \ce{^62Cu}, \ce{^64Cu}, and \ce{^89Zr} radioisotopes can be produced simultaneously when the incident photon energy is approximately 17 \si{MeV}.

Currently, \ce{^99Mo} is primarily produced by the $(n,f)$ reaction in a high-flux reactor using an enriched \ce{^235U} target. The specific activity of \ce{^99Mo} can reach 185 \si{TBq/g} (5000 \si{Ci/g}) ~\cite{b6}. However, most reactors will gradually shut down by 2030~\cite{b68}. Medical cyclotrons can be used to directly generate \ce{^{99m}Tc} via the \ce{^100Mo}$(p,2n)$ reaction. According to the experimental data, an enriched \ce{^100Mo} target irradiated with a 16.5 \si{MeV} proton beam at 130 \si{µA} for 6 \si{h} yielded 116 \si{GBq/g} (3.13 \si{Ci/g}) of \ce{^{99m}Tc}~\cite{b69}. Recently, a subcritical \ce{^99Mo} production system was developed, which is driven by an accelerator-based deuterium–deuterium (D–D) neutron source. The D–D fusion reaction generates neutrons that irradiate a low-enriched uranium solution and induce fission in \ce{^235U}. The system can generate 47.8 \si{mCi/g} \ce{^99Mo} for a stable 24 \si{h} operation with a neutron intensity of 1$\times10^{14}$ \si{n/s}~\cite{b70}. Bremsstrahlung gamma sources based on linear electron accelerators are often used to produce \ce{^99Mo} radioisotopes. Using a \ce{^100MoO_3} target irradiate with a 35 \si{MeV} electron beam at 100 \si{µA} for 20 \si{h}; the specific activity can achieve 4.4 \si{GBq/g} (119 \si{mCi/g})~\cite{b71}. NorthStar Medical Radioisotopes developed a \ce{^99Mo} production system using an electron linear accelerator. Using this system based on $(\upgamma,n)$ reactions, approximately 30\% more \ce{^99Mo} is produced per gram of the target material compared with the traditional neutron capture route~\cite{b72,b73}.

In comparison, the specific activity produced by the photonuclear method based on SLEGS was low. With a further increase in the flux of $\upgamma$ beams, from $10^7$ \si{s^{-1}} to $10^{15}$ \si{s^{-1}}~\cite{b20}, the specific activity of \ce{^99Mo} can reach up to 4$\times10^{2}$ \si{Ci/g}. For the 2-day protocol, the activity of the \ce{^{99m}Tc}-labelled tracers required for one myocardial perfusion imaging was 24 \si{mCi}~\cite{b74}. The total yield of stacked targets in one year can provide 400,000 myocardial perfusion images. As the flux of SLEGS increases in the future, this method is promising for producing medical radioisotopes and will become feasible in China.

\section*{Appendix}
\appendix
Experimental data for the production of medical radioisotopes by $(\upgamma,n)$, $(\upgamma,p)$, and $(\upgamma,\upgamma')$ reactions are summarized. Table~\ref{table5} lists relevant information on these reactions. All the data and information were obtained from the EXFOR database~\cite{b32}.

\begin{table*}[!htb]
\caption{Information regarding experimental data for the production of medical radioisotopes by photonuclear reactions}
\label{table5}
\begin{threeparttable}
\begin{tabular*}{\linewidth} {@{\extracolsep{\fill} } cccccc}
\toprule
Target             &         Reaction         &      Product     &  Year  &           Gamma source            &  Reference   \\
\midrule
\ce{^12C}\tnote{*} &      $(\upgamma,n)$      &     \ce{^11C}    &  1951  &          Bremsstrahlung           &  \cite{b55}  \\
                   &                          &                  &  1957  &          Bremsstrahlung           &  \cite{b75}  \\
                   &                          &                  &  1959  &  \ce{^3H}$(p,\upgamma)$\ce{^4He}  &  \cite{b76}  \\
                   &                          &                  &  1961  &          Bremsstrahlung           &  \cite{b77}  \\
                   &                          &                  &  1962  &  \ce{^3H}$(p,\upgamma)$\ce{^4He}  &  \cite{b78}  \\
                   &                          &                  &  1966  &          Bremsstrahlung           &  \cite{b79}  \\
                   &                          &                  &  1966  &          Bremsstrahlung           &  \cite{b80}  \\
                   &                          &                  &  1966  &  \ce{^3H}$(p,\upgamma)$\ce{^4He}  &  \cite{b81}  \\
\ce{^14N}          &      $(\upgamma,n)$      &     \ce{^13N}    &  1960  &          Bremsstrahlung           &  \cite{b82}  \\
                   &                          &                  &  1987  &          Bremsstrahlung           &  \cite{b83}  \\
\ce{^16O}\tnote{*} &      $(\upgamma,n)$      &     \ce{^15O}    &  1962  &  \ce{^3H}$(p,\upgamma)$\ce{^4He}  &  \cite{b78}  \\
                   &                          &                  &  1966  &          Bremsstrahlung           &  \cite{b84}  \\
                   &                          &                  &  1970  &          Bremsstrahlung           &  \cite{b85}  \\
                   &                          &                  &  1985  &          Bremsstrahlung           &  \cite{b86}  \\
                   &                          &                  &  1991  &          Bremsstrahlung           &  \cite{b87}  \\
\ce{^19F}          &      $(\upgamma,n)$      &     \ce{^18F}    &  1962  &  \ce{^3H}$(p,\upgamma)$\ce{^4He}  &  \cite{b78}  \\
\ce{^63Cu}         &      $(\upgamma,n)$      &     \ce{^62Cu}   &  1951  &          Bremsstrahlung           &  \cite{b88}  \\
                   &                          &                  &  1951  &          Bremsstrahlung           &  \cite{b55}  \\
                   &                          &                  &  1954  &          Bremsstrahlung           &  \cite{b89}  \\
                   &                          &                  &  1955  &          Bremsstrahlung           &  \cite{b90}  \\
                   &                          &                  &  1959  &  \ce{^7Li}$(p,\upgamma)$\ce{^8Be} &  \cite{b91}  \\
                   &                          &                  &  1960  &  \ce{^7Li}$(p,\upgamma)$\ce{^8Be} &  \cite{b92}  \\
                   &                          &                  &  1961  &  \ce{^7Li}$(p,\upgamma)$\ce{^8Be} &  \cite{b57}  \\
                   &                          &                  &  1962  &  \ce{^3H}$(p,\upgamma)$\ce{^4He}  &  \cite{b78}  \\
                   &                          &                  &  1968  &  In-flight positron annihilation  &  \cite{b93}  \\
                   &                          &                  &  1968  &          Bremsstrahlung           &  \cite{b94}  \\
                   &                          &                  &  1979  &  In-flight positron annihilation  &  \cite{b95}  \\
                   &                          &                  &  1984  &  \ce{^7Li}$(p,\upgamma)$\ce{^8Be} &  \cite{b96}  \\
                   &                          &                  &  1984  &          Bremsstrahlung           &  \cite{b97}  \\
                   &                          &                  &  2012  &          Bremsstrahlung           &  \cite{b98}  \\
\ce{^65Cu}\tnote{*}&      $(\upgamma,n)$      &     \ce{^64Cu}   &  1951  &          Bremsstrahlung           &  \cite{b55}  \\
                   &                          &                  &  1961  &  \ce{^7Li}$(p,\upgamma)$\ce{^8Be} &  \cite{b57}  \\
                   &                          &                  &  1990  &          Bremsstrahlung           &  \cite{b56}  \\
\ce{^90Zr}         &      $(\upgamma,n)$      &     \ce{^89Zr}   &  1978  &          Bremsstrahlung           &  \cite{b99}  \\
                   &                          &                  &  2010  &          Bremsstrahlung           &  \cite{b100}  \\
                   &                          &                  &  2019  &               LCS                 &  \cite{b101}  \\
\ce{^100Mo}        &      $(\upgamma,n)$      &     \ce{^99Mo}   &  1978  &          Bremsstrahlung           &  \cite{b99}  \\
                   &                          &                  &  2011  &          Bremsstrahlung           &  \cite{b49}  \\
                   &                          &                  &  2011  &               LCS                 &  \cite{b14}  \\
                   &                          &                  &  2013  &               LCS                 &  \cite{b48}  \\
\ce{^187Re}        &      $(\upgamma,n)$      &    \ce{^186Re}   &  2005  &               LCS                 &  \cite{b102}  \\
\ce{^44Ca}         &      $(\upgamma,p)$      &     \ce{^43K}    &  1978  &          Bremsstrahlung           &  \cite{b99}  \\
\ce{^68Zn}         &      $(\upgamma,p)$      &    \ce{^67Cu}    &  1957  &          Bremsstrahlung           &  \cite{b67}  \\
                   &                          &                  &  2021  &          Bremsstrahlung           &  \cite{b56}  \\
\ce{^178Hf}        &      $(\upgamma,p)$      &   \ce{^177Lu}    &  2020  &          Bremsstrahlung           &  \cite{b103}  \\
\ce{^103Rh}        &  $(\upgamma,\upgamma')$  &  \ce{^{103m}Rh}  &  1961  &          Bremsstrahlung           &  \cite{b104}  \\
\ce{^113In}        &  $(\upgamma,\upgamma')$  &  \ce{^{113m}In}  &  2006  &          Bremsstrahlung           &  \cite{b105}  \\
                   &                          &                  &  2017  &          Bremsstrahlung           &  \cite{b106}  \\
\ce{^115In}        &  $(\upgamma,\upgamma')$  &  \ce{^{115m}In}  &  1957  &          Bremsstrahlung           &  \cite{b107}  \\
                   &                          &                  &  1993  &          Bremsstrahlung           &  \cite{b108}  \\
                   &                          &                  &  2001  &          Bremsstrahlung           &  \cite{b109}  \\
                   &                          &                  &  2006  &          Bremsstrahlung           &  \cite{b110} \\
                   &                          &                  &  2017  &          Bremsstrahlung           &  \cite{b106}  \\
                   &                          &                  &  2018  &               LCS                 &  \cite{b111} \\
\ce{^195Pt}        &  $(\upgamma,\upgamma')$  &  \ce{^{195m}Pt}  &  2006  &          Bremsstrahlung           &  \cite{b105}  \\
\bottomrule
\end{tabular*}
\begin{tablenotes}
\footnotesize
    \item[*] Experimental data within a specific energy range, instead of all of experimental data
\end{tablenotes}
\end{threeparttable}
\end{table*}
\clearpage

\subsection{$(\upgamma,n)$}
The cross-sectional data for the production of \ce{^11C}, \ce{^13N}, \ce{^15O}, \ce{^18F}, \ce{^62Cu}, \ce{^64Cu}, \ce{^89Zr}, \ce{^99Mo}, and \ce{^186Re} radioisotopes by $(\upgamma,n)$ reactions are shown in the following figures(Figs.~\ref{fig4}, ~\ref{fig5}, ~\ref{fig6}, ~\ref{fig7}, ~\ref{fig8}, ~\ref{fig9}, ~\ref{fig10}, ~\ref{fig11} and ~\ref{fig12}).

\begin{figure}[H]
\centering
\includegraphics[width=0.7\hsize]{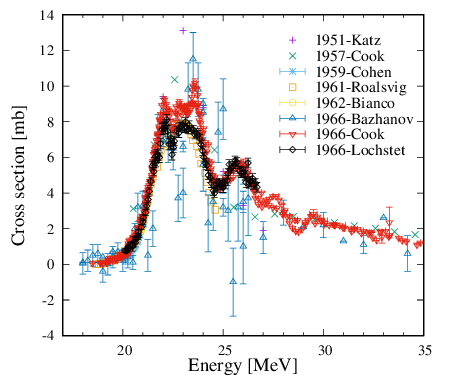}
\caption{(Color online) \ce{^12C}$(\upgamma,n)$\ce{^11C} reaction cross sections.}
\label{fig4}
\end{figure}

\begin{figure}[H]
\centering
\includegraphics[width=0.7\hsize]{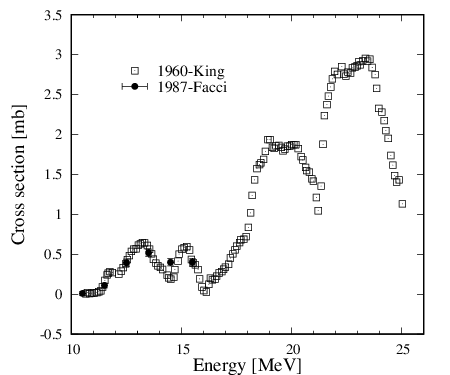}
\caption{\ce{^14N}$(\upgamma,n)$\ce{^13N} reaction cross sections.}
\label{fig5}
\end{figure}

\begin{figure}[H]
\centering
\includegraphics[width=0.7\hsize]{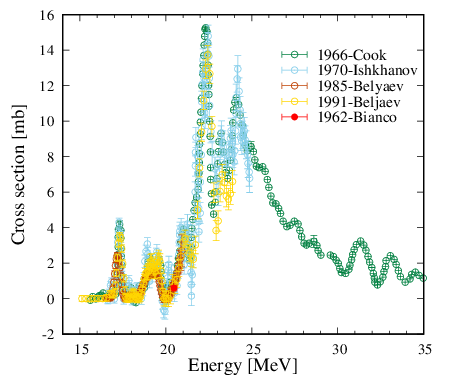}
\caption{(Color online) \ce{^16O}$(\upgamma,n)$\ce{^15O} reaction cross sections.}
\label{fig6}
\end{figure}

\begin{figure}[H]
\centering
\includegraphics[width=0.7\hsize]{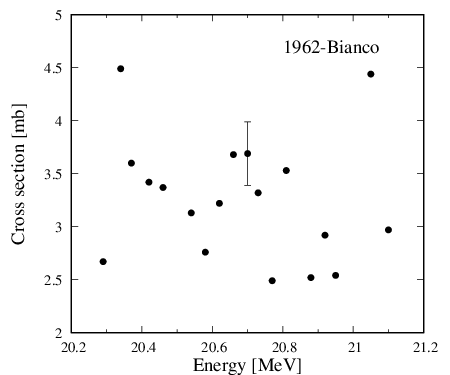}
\caption{\ce{^19F}$(\upgamma,n)$\ce{^18F} reaction cross sections.}
\label{fig7}
\end{figure}

\begin{figure}[H]
\centering
\includegraphics[width=0.7\hsize]{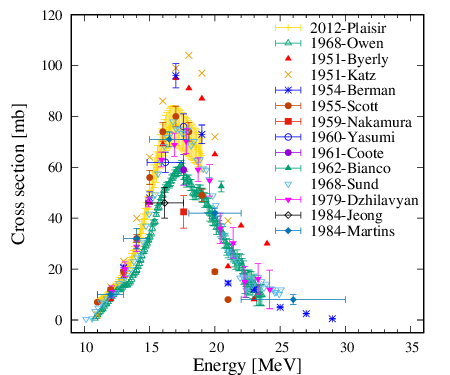}
\caption{(Color online) \ce{^63Cu}$(\upgamma,n)$\ce{^62Cu} reaction cross sections.}
\label{fig8}
\end{figure}

\begin{figure}[H]
\centering
\includegraphics[width=0.7\hsize]{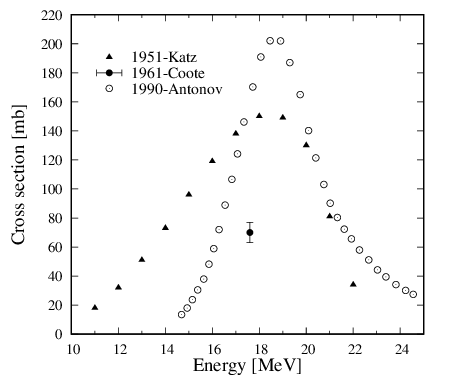}
\caption{\ce{^65Cu}$(\upgamma,n)$\ce{^64Cu} reaction cross sections.}
\label{fig9}
\end{figure}

\begin{figure}[H]
\centering
\includegraphics[width=0.7\hsize]{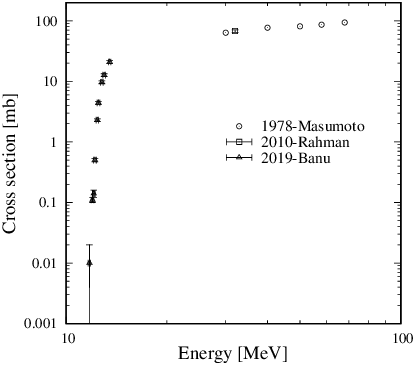}
\caption{\ce{^90Zr}$(\upgamma,n)$\ce{^89Zr} reaction cross sections.}
\label{fig10}
\end{figure}

\begin{figure}[H]
\centering
\includegraphics[width=0.7\hsize]{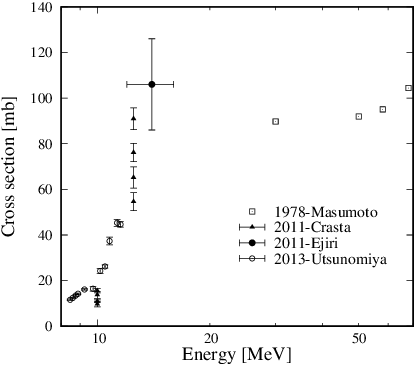}
\caption{\ce{^100Mo}$(\upgamma,n)$\ce{^99Mo} reaction cross sections.}
\label{fig11}
\end{figure}

\begin{figure}[H]
\centering
\includegraphics[width=0.7\hsize]{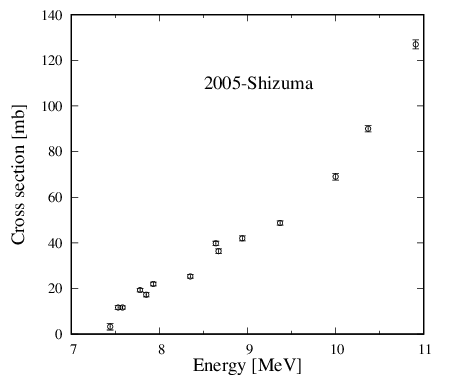}
\caption{\ce{^187Re}$(\upgamma,n)$\ce{^186Re} reaction cross sections.}
\label{fig12}
\end{figure}

\subsection{$(\upgamma,p)$}
The cross-sectional data for the production of \ce{^43K}, \ce{^67Cu}, and \ce{^177Lu} radioisotopes by $(\upgamma,p)$ reaction are shown in the following figures(Figs.~\ref{fig13}, ~\ref{fig14} and ~\ref{fig15}).

\begin{figure}[H]
\centering
\includegraphics[width=0.7\hsize]{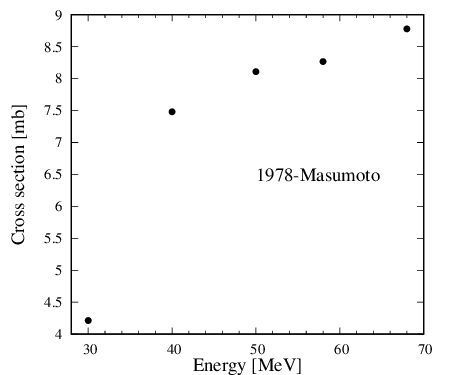}
\caption{\ce{^44Ca}$(\upgamma,p)$\ce{^43K} reaction cross sections.}
\label{fig13}
\end{figure}

\begin{figure}[H]
\centering
\includegraphics[width=0.7\hsize]{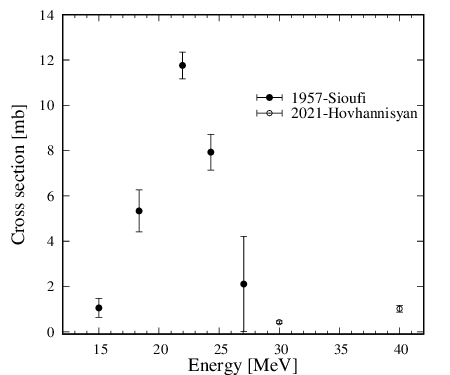}
\caption{\ce{^68Zn}$(\upgamma,p)$\ce{^67Cu} reaction cross sections.}
\label{fig14}
\end{figure}

\begin{figure}[H]
\centering
\includegraphics[width=0.7\hsize]{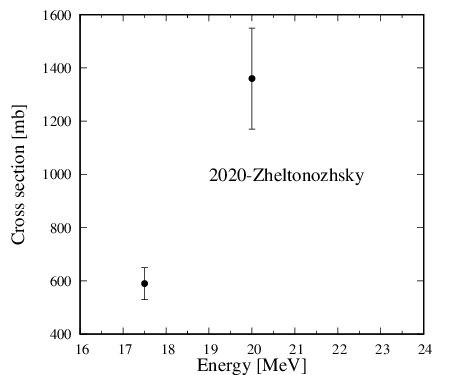}
\caption{\ce{^178Hf}$(\upgamma,p)$\ce{^177Lu} reaction cross sections.}
\label{fig15}
\end{figure}

\subsection{$(\upgamma,\upgamma')$}
The cross-sectional data for the production of \ce{^{103m}Rh}, \ce{^{113m}In}, \ce{^{115m}In}, and \ce{^{195m}Pt} radioisotopes by $(\upgamma,\upgamma')$ reaction are shown in the following figures(Figs.~\ref{fig16}, ~\ref{fig17}, ~\ref{fig18} and ~\ref{fig19}).

\begin{figure}[H]
\centering
\includegraphics[width=0.7\hsize]{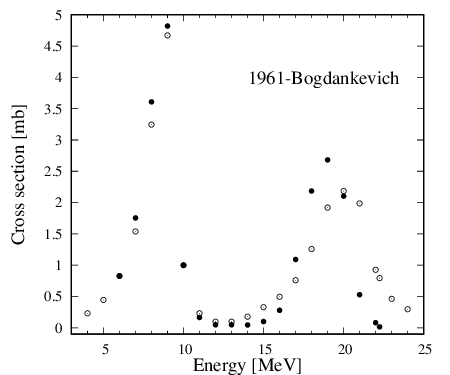}
\caption{\ce{^103Rh}$(\upgamma,\upgamma')$\ce{^{103m}Rh} reaction cross sections.}
\label{fig16}
\end{figure}

\begin{figure}[H]
\centering
\includegraphics[width=0.7\hsize]{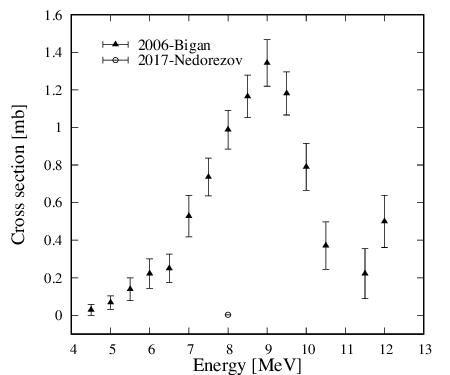}
\caption{\ce{^113In}$(\upgamma,\upgamma')$\ce{^{113m}In} reaction cross sections.}
\label{fig17}
\end{figure}

\begin{figure}[H]
\centering
\includegraphics[width=0.7\hsize]{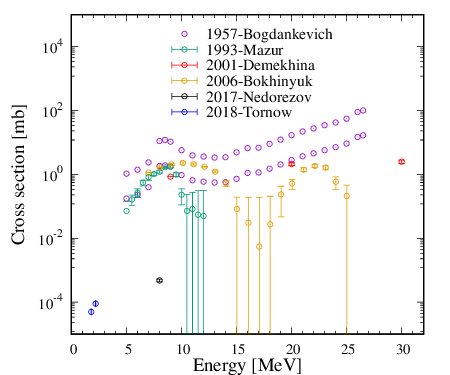}
\caption{(Color online) \ce{^115In}$(\upgamma,\upgamma')$\ce{^{115m}In} reaction cross sections.}
\label{fig18}
\end{figure}

\begin{figure}[H]
\centering
\includegraphics[width=0.7\hsize]{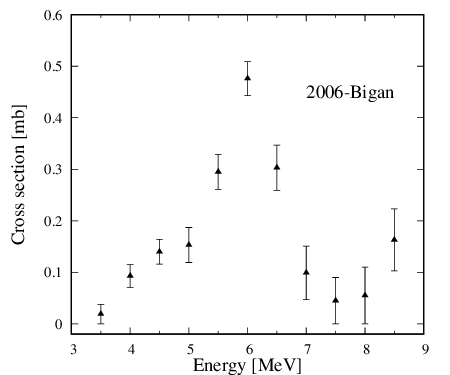}
\caption{\ce{^195Pt}$(\upgamma,\upgamma')$\ce{^{195m}Pt} reaction cross sections.}
\label{fig19}
\end{figure}

\end{document}